\documentclass[preprint,12pt]{elsarticle}

\usepackage{amssymb} 

\journal{Forces in Mechanics}

\begin{document}

\begin{frontmatter}



\title{Predicting Creep Failure by Machine Learning - Which Features Matter?}

\author
{Stefan Hiemer,$^{1}$, Paolo Moretti$^{1}$, 
Stefano Zapperi,$^{1,2,3}$, Michael Zaiser$^{1\ast}$\corref{cor}\\
\normalsize{$^{1}$ Institute of Materials Simulation, Department of Materials Science Science and Engineering,}\\
\normalsize{Friedrich-Alexander-University Erlangen-Nuremberg, 
Dr.-Mack-Str. 77, 90762 F{\"u}rth, Germany}\\
\normalsize{$^{2}$ Center for Complexity and Biosystems, Department of Physics,}\\
\normalsize{University of Milan, via Celoria 16, 20133 Milan, Italy}\\
\normalsize{$^{3}$ CNR - Consiglio Nazionale delle Ricerche,}\\ 
\normalsize{Istituto di Chimica della Materia Condensata e di Tecnologie per l'Energia,}\\
\normalsize{Via R. Cozzi 53, 20125 Milan, Italy}\\
\normalsize{$^\ast$Corresponding author; E-mail:  michael.zaiser@fau.de}
}
            
\begin{abstract}
Spatial and temporal features are studied with respect to their predictive value for failure time prediction in subcritical failure with machine learning (ML). Data are generated from simulations of a novel, brittle random fuse model (RFM), as well as elasto-plastic finite element simulations (FEM) of a stochastic plasticity model with damage, both models considering stochastic thermally activated damage/failure processes in disordered materials. Fuse networks are generated with hierarchical and nonhierarchical architectures. Random forests - a specific ML algorithm - allow us to measure the feature importance through a feature's average error reduction. RFM simulation data are found to become more predictable with increasing system size and  temperature. Increasing the load or the scatter in local materials properties has the opposite effect. Damage accumulation in these models proceeds in stochastic avalanches, and statistical signatures such as avalanche  rate or magnitude have been discussed in the literature as predictors of incipient failure. However, in the present study such features proved of no measurable use to the ML models, which mostly rely on global or local strain for prediction. This suggests the strain as viable quantity to monitor in future experimental studies as it is accessible via digital image correlation.
\end{abstract} 


\end{frontmatter}


\section{Introduction}

Creep failure is an example of a subcritical failure process, where an applied load which is insufficient to instantaneously break the sample drives time dependent damage accumulation. This gradual accumulation of damage deteriorates the strength of the material and ultimately results in delayed failure \cite{andrade1910viscous}. 

It is in general unfeasible to design structures in such a manner as to avoid the damage processes that lead to subcritical failure. Predicting the residual lifetime of a structure under subcritical load is therefore an important issue that is actively investigated by both physicists and engineers. Reliable lifetime predictions may help to avoid catastrophic in-service failure of components and systems and harness substantial economic benefits by adapting and where possible extending replacement cycles. 

To assist prediction, it is desirable to obtain sample specific information about the damage accumulation process through non destructive means. Such information can be obtained from the macroscopic sample response, i.e., the time dependent creep strain or strain rate as accessible by surface monitoring. Additional and more detailed information can be drawn from analysis of the spatio-temporal pattern of energy releases as local creep damage accumulates, as microcrack formation is accompanied by elastic energy release which can be recorded by monitoring the acoustic emission (AE) of the sample.

Several empirical approaches have been proposed to predict sample specific failure times from macroscopic creep strain data. The simplest possible approach is to correlate the time $t_{\rm m}$ of minimum strain rate with the catastrophic failure time $t_{\rm f}$, in the simplest case by assuming a linear relationship between both \cite{hao2014predicting,koivisto2016predicting}. A variant consists in relating the failure time to the duration of the primary (decelerating) creep stage \cite{nechad2005creep}. 

A slightly different approach towards failure time prediction based on macroscopic strain (strain rate) focuses, instead, on the rapid increase of creep strain and strain rate in the run-up to failure, which typically is characterized by a creep strain (strain rate) that increases like an inverse power of the time-to-failure. Fitting such a to the data recorded until a given moment implies a prediction of the residual lifetime -- an approach which has been promoted by D. Sornette and applied, in different variations, to catastrophic phenomena from material rupture over financial crises to childbirth \cite{sornette2002predictability} to the catastrophic breakdown of civilization as we know it \cite{johansen2001finite}. 

A related prediction approach focuses on temporal statistics of precursor events, whose magnitudes and rates also may develop characteristic singularities in the approach to failure. For instance, one may exploit the observation made both in simulations \cite{castellanos2018} and experiments \cite{lennartz2014acceleration} that the AE event rate $\nu_{\rm AE}$ accelerates towards failure according to a reverse Omori law, $\nu_{\rm AE} \propto (t - t_{\rm f})^{-p}$ with $p \approx 1$. This behavior is also found in mean-field models of thermally activated rupture \cite{saichev2005andrade} and allows to obtain the failure time by fitting the Omori law to the previous AE record. At the same time, the approach to failure may be accompanied with other characteristic changes in the AE burst statistics, such as an increase in the AE event size or characteristic changes in the Gutenberg-Richter exponent of the power law type energy statistics \cite{castellanos2018}, which may also be used for monitoring and prediction. 

Beyond the temporal record of strain, strain rate and AE, additional information can be obtained by simultaneously monitoring the spatial pattern of local strains, or more generally of damage accumulation. Materials failure is associated with localization of damage \cite{lennartz2014acceleration,castellanos2018}, and signatures of damage or strain localization may provide additional features that assist failure forecasting. In the present investigation, we use machine learning (ML) approaches to assess the relative usefulness of spatial and temporal features of damage accumulation in view of the forecasting of creep failure times. Machine learning has been used in the context of fracture and failure to identify critical conditions for load driven failure together with crack nucleation sites and propagation pathways in amorphous silica \cite{font2022predicting}. In the context of subcritical failure, ML has been used to predict sample specific failure times in disordered solids by means of Random Forest regression \cite{biswas}, and similar methodology was used to forecast laboratory earthquakes from AE time series records \cite{rouet2017machine}. In the latter context, \cite{johnson2021laboratory} have issue a Kaggle challenge to benchmark the performance of different ML algorithms. 

In the present study, we use ML to assess the usefulness of different spatial and temporal features for predicting creep failure times. As data base we consider simulation data from two different types of models, namely a highly simplified model (random fuse model, RFM) of thermally activated damage accumulation, as well as the stochastic FEM model of \cite{castellanos2018} that was used in the previous work by Biswas et. al. \cite{biswas}. These models are introduced in Section 2 together with the respective feature sets extracted from the simulations for sample specific failure time prediction. The behavior of the RFM in the run-up to failure is discussed in Section 3.1, while results of the ML analysis of the data sets are given in Section 3.2. Section 4 concludes with a critical discussion of the usefulness of different features, and the possibilities and limitations of sample specific lifetime prediction. 

\section{Methods}

\subsection{RFM simulations}
\begin{figure}[tb]
\includegraphics[width=\textwidth]{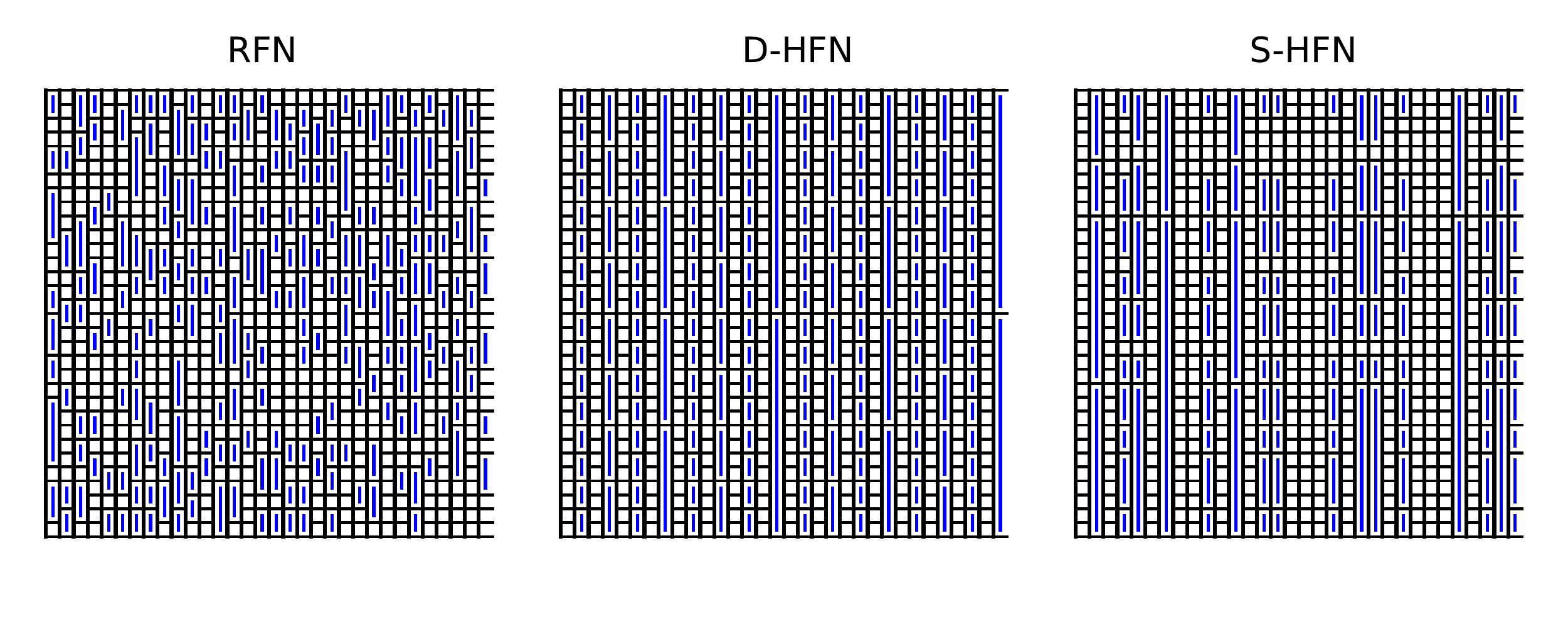}
\caption{Two-dimensional fuse networks of size $L=32$. Eternal uni-axial loads are applied at the the top and bottom boundaries. All systems share the same number of horizontal cross-linking fuses. Gaps are emphasized in blue. Gap sizes are exponentially distributed in the non-hierarchical case (RFN) and power-law distributed in the hierarchical case (D- and S-HFN)}
\label{fig:schematic}
\end{figure}

We consider simplifications failure, so called random fuse models, which provide a scalar caricature of elastic-brittle behavior by modelling materials as networks of scalar load carrying elements. We envisage two-dimensional structures as shown in Figure \ref{fig:schematic}. In mechanical terms, these structures can be envisaged as modelling sheet-like materials deformed in plane stress conditions. The structures are constituted of beam-like load carrying elements of unit length, which form a two-dimensional lattice of junction points, or {\it nodes}. A tensile-like load is applied in the vertical direction. In the continuum limit of infinitely many load carrying elements, the scalar character of load and displacement corresponds to a material of zero Poisson ratio zero loaded in uni-axial tension, for which the equations of elastostatics reduce to the Laplace equation. 

The elements are arranged using the following architecture: For a given system size $L$, $L^2$ vertically oriented elements transmit load across the system in the load-parallel direction, while a fixed number $C<L$ of horizontal cross-linking elements is responsible for load redistribution in the load-perpendicular direction. Figure \ref{fig:schematic} shows three possible variants of this construction. In the Random Fuse Network (RFN), horizontal cross-linking elements are distributed randomly. This arrangement results in the formation of vertical {\it gaps}, which interrupt load redistribution and exhibit an exponential length distribution \cite{moretti2018avalanche}. A deterministic hierarchical fuse network (D-HFN) refers to a similar system, where however the cross links are distributed hierarchically, and the resulting gaps have a heavy tailed, power-law size distribution \cite{moretti2018avalanche}. Finally, a shuffled hierarchical fuse network (S-HFN) is constructed from a D-HFN, where rows and columns of the network adjacency matrix are randomly shuffled. This construction maintains the power-law gap-size distribution  and captures the same failure phenomenology of the D-HFN, but at the same time allows for averaging across different network realizations. In all systems, periodic boundary conditions are imposed in the load perpendicular direction. In the actual simulations a D-HFN pattern for a prescribed size $L$ is identified first, and RFN and S-HFN variants are generated from that, making sure that the number of horizontal fuses is the same. In the following, we restrict our study to RFN and S-HFN, and we refer to S-HFN simply as HFN from now on. 

The dynamics of damage accumulation and fracture is best described using an electrical analogue, hence the term 'random fuse model' \cite{deArcangelis1985_JPL,Zapperi1999_PRE}. Owing to the scalar nature of the load variable, the load carrying elements can be envisaged as fuses carrying currents according to scalar constitutive equations. Identifying a fuse $ij$ by the indices of its two end-nodes $i$ and $j$, we use $I_{ij}$ to indicate the scalar force (current) acting on $ij$, and we relate it to the scalar displacements (voltages) $V_i$ and $V_j$ of the fuse endpoints through the scalar Hooke's law (Ohm's law)
\begin{equation}\label{eq:hooke}
I_{ij} = V_i-V_j,
\end{equation}
where we have assumed a unit elastic modulus (Resistance), while the balance equation for each node $i$ can be computed imposing that the algebraic sum of the scalar forces acting on $i$ is zero (in the electrical analogue: Kirchhoff's law). Externally imposed displacements are simulated by applying prescribed voltages at the boundaries. The response of the system is then captured by monitoring the applied voltage $V$ and the global current $I$.  The RFM implies a simplification of the elastic problem, which can be physically realized by plane stress loading of a sheet of zero Poisson ratio. Despite the simplifications, essential aspects of fracture mechanics such as load re-distribution and stress singularities at crack tips are preserved in the continuum limit of a RFM, and RFM models have for this reason been widely adopted in statistical modelling of fracture \cite{alava2006}. 
 
In order to account for local strength fluctuations, each fuse $ij$ is assigned a threshold $\theta_{ij}$, corresponding to a critical current at which it fails irreversibly (i.e., the fuse conductivity is set to zero). Thresholds are distributed following a Weibull cumulative distribution function of the form
\begin{equation}
C(t_{ij}) = 1 - \exp\left[-\left( \frac{\theta_{ij}}{\theta_0}  \right)^k    \right],
\end{equation}
where we choose $\theta_0=1/\Gamma(1+1/k)$ to set the mean value $\langle t_{ij}\rangle = 1$, and we vary the shape parameter $k$ to control the degree of heterogeneity in the material model (lower $k$ leading to larger fluctuations)  \cite{weibull}.  

To simulate sub-critical, thermally activated failure, we vary voltages at the boundaries so to ensure that the total current $I$ is constant. Fuses may in this model fail either instantaneously by overloading ($\theta_{ij}-I_{ij} < 0$) or by thermal activation, which is considered in terms of crossing an energy barrier
\begin{equation}
    \Delta U = \theta_{ij}-I_{ij}.
\end{equation}
 Whenever $\theta_{ij}\le I_{ij}$ holds for any number of fuses, the corresponding fuses $ij$ are removed instantaneously, equilibrium equations are solved again and the process is repeated until no instantaneous removals occur. When $theta_{ij} > I_{ij}$ for every fuse $ij$, a Kinetic Monte Carlo step is performed to identify the fuse which fails next because of thermally assisted energy barrier crossing, and to determine the time interval $\Delta t$ after which this happens. The attempts to cross barriers are assumed to be stochastically independent, and thus described by Poisson  processes with the transition rates
\begin{equation}
    \nu_{ij}=\nu_{0}\exp\left(-\frac{\theta_{ij}-I_{ij}}{T}\right)
    \label{eq_rate}
\end{equation}
with the characteristic frequency \(\nu_{0}\) and the temperature \(T\)  scaled with an appropriate constant of (e. g. Boltzmann constant). The unit of time of the simulations is \(\nu_{0}^{-1}\), thus we set \(\nu_{0}=1\) without loss of generality. A failing fuse is selected and removed accordingly, and time is increased by an interval $\Delta \tau$ extracted from an exponential distribution with mean value $\nu^{-1}$ where the total event rate is
\begin{equation}
    \nu = \sum_{ij\in \mathcal{F}}\nu_{ij}.
\end{equation}
Here \(\mathcal{F}\) refers to the set of surviving fuses at a given time. Note that thermally activated failure of one fuse may, due to load re-distribution, lead to overloading of other fuses which then fail instantaneously. In this case we speak of an {\em avalanche} of size $s$ where $s$ is the number of fuses that fail in a correlated manner until the stability condition $\theta_{ij} > I_{ij} \forall _{ij}$ is met. The next event is then again thermally activated. 

A subset of simulations is also run resorting to an extremal (i.e. non-stochastic) failure criterion (see e.g. Alava {\it et al.} for details \cite{alava2006}) which is equivalent  to the zero-temperature limit of the protocol outlined above.In extremal simulations, due to the absence of thermal activation all fuses fail by overloading -- either by an increase in applied load or at constant load in an avalanche. Damage accumulation and failure are driven by increasing the applied load, and extremal simulations thus allow us to identify the global peak load (peak current) $I_\mathrm{p}$ which can be supported by the system of fuses before it undergoes instantaneous failure. 

Subcritical creep loading, on the other hand, is performed by applying a constant load $I$ in the interval $]0,I_\mathrm{p}[$ at finite temperature, such that fuses can fail by thermal activation. For all considered combinations of parameter values (Tab. \ref{random-fuse-parameter-space}) we gathered data from multiple samples, performing for each parameter combination 1200 simulations with different sets of random thresholds.  

\begin{table}[ht]
\label{random-fuse-parameter-space}
\caption{Choice of parameters investigated for the RFM. All combinations of these parameters are investigated with machine learning except current 0.1 for size 128.}
\center
\begin{tabular}{cc}
\hline
parameter &  parameter values \\
\hline
\(k\) & 1,2,4,8,16 \\
\(T\) & 0.01,0.03,0.05,0.1 \\
\(I/I_{p}\) & (0.1),0.3,0.5,0.7,0.9 \\
\(L\) & 16,32,64,128 \\
\hline
\end{tabular} 
\end{table}

\subsection{Finite Element Data}

As an alternative source of information, data created in a previous study \cite{biswas} from a two-dimensional elasto-plastic creep finite element (FEM) model, considering a 2D square block that is loaded in simple plane shear, have been used. For details of this simulation we refer to \cite{castellanos2018,castellanos2019}. The FEM and RFM creep models have some differences. i) The RFM model is equivalent to a periodically continued domain loaded in pure anti-plane shear, where the out-of-plane shear stress corresponds to the current of the electrical analogue. The FEM model, on the other hand, considers plane strain deformation which makes a tensorial treatment mandatory. The current is now replaced by the von Mises equivalent stress calculated from the deviatoric stress, the local threshold is the counterpart of a local yield stress, and the role of fuses is taken over by the elements. ii) Once the threshold is exceeded by the stress, the element deforms by a plastic incremental strain $\Delta \epsilon$. Unlike the RFM, elements can deform repeatedly, however, with each threshold crossing damage is added to the element. After the threshold crossing, a new threshold is re-drawn from a Weibull distribution with shape parameter \(k\) whose mean value decreases with the accumulated element damage. Again, deformation proceeds either by thermally activated events or by stress re-distribution leading to overloading, causing avalanches. The avalanche size is here defined as the number of strain increments caused by correlated overloading between two consecutive thermally activated barrier crossing events. The simulation ends once the system enters a never-ending avalanche, pragmatically defined as an avalanche size that exceeds the total number of elements. \\

The creep load is again measured in units of the zero-temperature failure strain of the specific system. For stress 0.7, \(k\,\in\,\{1,2,4,8,16\}\) is investigated while for stress 0.9 just \(k=4\). The data recorded consist of the times and locations of thermally activated events as well as the sizes of the ensuing avalanches.

\subsection{Random Forests and Feature Selection}
Random forests are constructed by bootstrapping of decision trees. In the context of machine learning, bootstrapping takes the average over the predictions of an ensemble of regressors that have been trained on different random subsamples of the training data. This mitigates against overfitting (for which single decision trees are notorious) and provides additional probabilistic information in terms of the statistical distribution of predictions. Single decision trees divide the feature space recursively into partitions. The predicted value of a variable -- here: the residual lifetime -- in each partition is the average of the actual values of that variable´for the training samples that fall into the partition. The partitioning is done on form of binary decisions ('splits') based on single features, taking in each recursive step the decision which minimizes the mean squared error of the ensuing predictions. Feature importance is determined in terms of the (normalized) average error reduction that is achieved by making splits based on a given feature  \cite{louppe2014understanding}. For details we refer to \cite{breimanforest,hoforest1995,hoforest1998}. 

Scikit version 0.24.2 is used for the machine learning \cite{scikit-learn}. Default parameters the random forest are used because preliminary studies as well as the previous work of Biswas et al \cite{biswas} indicated that the results to be insensitive to the hyperparameters of the random forest. To measure prediction performance, besides summary statistical signatures like the mean squared error, the machine learning prediction (ML) at time \(t\) is compared to a set of four simple baseline predictions \(t_{\rm BA}\)
\begin{equation}
    t_{\rm BA}(t) = (t_{\rm ref}-t)H(t_{\rm ref}-t)
\end{equation}
, where \(t_{\rm ref}\) is one of four reference times. \(H\) is the Haeviside function to avoid prediction of negative remaining time to failure. Two simple reference times are the average and median lifetime of the entire training set. For the other two, one compares \(t\) with the lifetimes \(t_{\rm f}\) in the training set, ignores all lifetimes smaller than \(t\) and takes the mean or median lifetime of the remaining samples. We refer to the first two reference as static mean and median whereas the last two as dynamic mean and median.

For the FEM data, two different feature sets have been used: The features described the previous work of Biswas et. al. \cite{biswas} and a new set consisting of time, total damage as proxy for the total strain, thermally activated event rate, the average avalanche amplitude over a finite time window, and a spatial damage localization parameter. The latter quantity is defined resorting to domain knowledge, which tells us that for the imposed loading conditions (simple shear imposed on the boundaries of a quadratic domain), an incipient shear band must be aligned with the domain boundaries. Accordingly, we determine the maximum value of strain (or equivalently damage) within a rectangle of predefined width that spans the system parallel to one of its boundaries, i.e. in one of the two directions of a potential shear band. The rectangle width is varied (2,4,8,32 elements) such as to achieve optimum prediction performance. To calculate the thermally activated event rate, a running average is used over  \(n_{\rm av}={\rm floor}(\frac{\bar{N}_{\rm av}}{X})\) events where \(\bar{N}_{\rm av}\) is the average number of thermally activated events in a simulation and \(X\) an integer value that needs to be chosen. We refer to \(X\) as rate parameter. In this study the values \(50,100,200\) are compared. The event rate \(r\) at  the \(i\)-th event then is defined as 
\begin{equation}
    r_{i}= \frac{n_{\rm av}}{(t_{i}-t_{(i-n_{\rm av})})}
\end{equation}
The average avalanche amplitude is calculated over the same running window.

For the RFM data, also two feature sets are used. For the first set, the features are time, voltage, the total damage measured as the number of failed fuses, as well as the maximum number of failed fuses aligned in the direction perpendicular to the applied load, which corresponds to the propagation  direction of an emergent tensile crack.  The second feature set consists of time, voltage, the total damage measured as the number of failed fuses and the maximum crack size. The latter is obtained by determining all connected clusters of broken fuses. The  size of such a cluster is defined as the number of broken fuses. The largest crack, i.e. the largest connected component, is found by using the graph implementation of scipy \cite{scipy}. 

\begin{table}[th]
\label{feature-sets-rfm-table}
\caption{Overview over feature sets used for the RFM data.}
\center
\begin{tabular}{cc}
\hline
 &  FEM features \\
\hline
Set 1 & time, avalanche size,min. line damage, max. line damage \\
Set 2 & time, avalanche rate, average avalanche amplitude \\
 & total damage \\
\hline
 &  RFM features \\
\hline
Set 1 & time, voltage, max. line damage vertical \\
 & max. line damage horizontal, total damage \\
Set 2 & time, voltage, total damage, max. crack size \\
\hline
\end{tabular} 
\end{table}

\section{Results and Discussion}
\subsection{Macroscopic RFM model behaviour}

\begin{figure}[t]
\centering
\includegraphics[width=.7\textwidth]{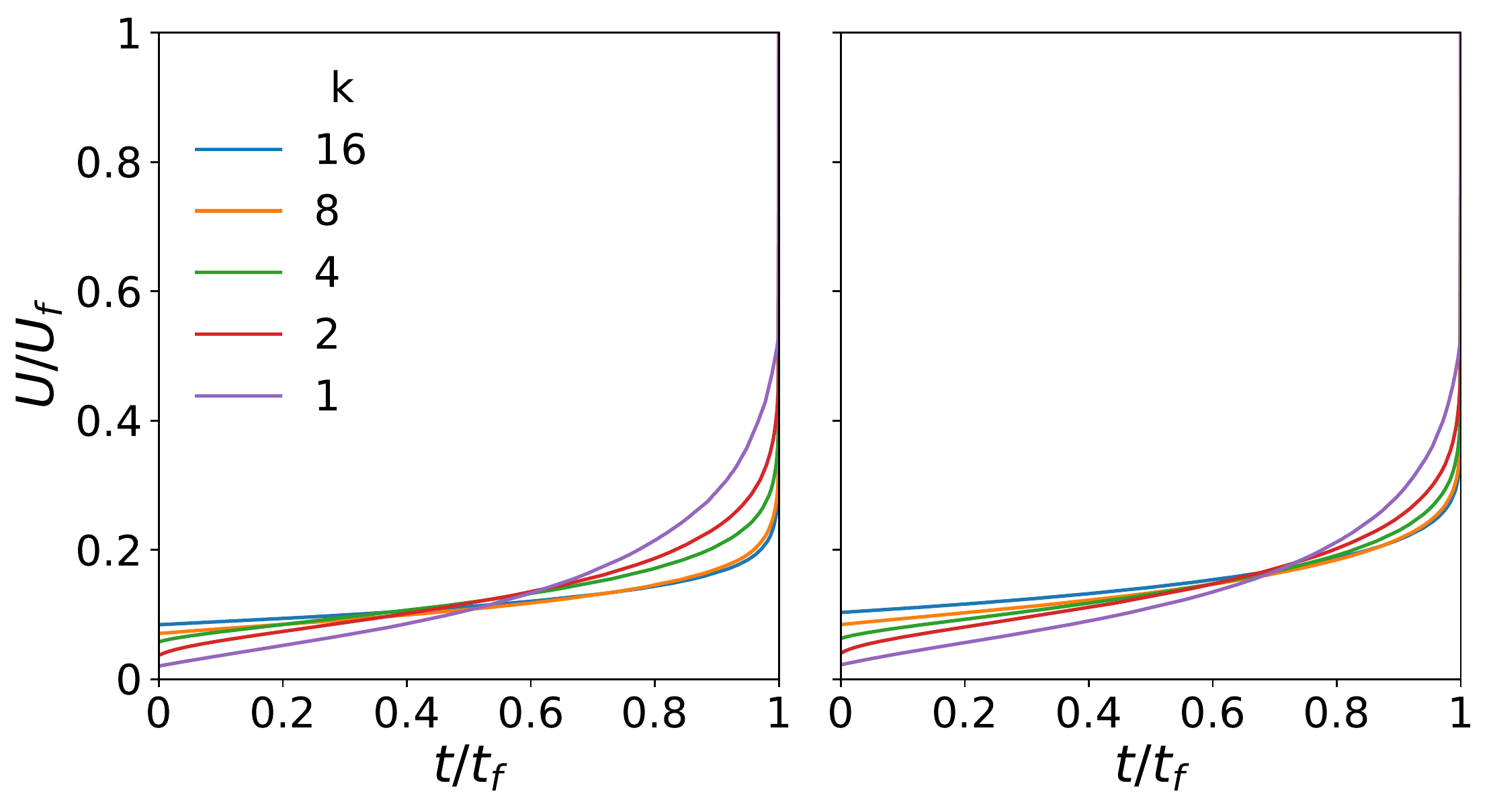}
\caption{Normalized average time-voltage curve for \(L=128\),\(\frac{I}{I_{p}} =0.1\) and \(T=0.1\). Data of nonhierarchical networks are on the right and of hierarchical networks on the left.}
\label{t-U-closeup}
\end{figure}

Figure \ref{t-U-closeup} shows average time-voltage curves, for a relative applied load $\sigma_0 = 0.1$ and temperature $T=0.1$, and for varying Weibull shape parameters $k$, in hierarchical and non-hierarchical systems of size $L=128$. In a typical creep strain vs time curve, one can identify three regimes: a strain hardening regime at small $t / t_\mathrm{f}$ during which the strain rate decreases, a stationary regime with almost constant strain rate for intermediate  $t / t_\mathrm{f} $, and strain softening towards the end of the sample lifetime \(t_\mathrm{f}\). Figure \ref{t-U-closeup} confirms this picture, with a deviation in the case of very low disorder (high $k$) where the initial strain hardening part becomes less and less prominent. A more complete picture of the influence of simulation parameters on the behavior of the system is presented in the Appendix (Figures S1 ,S2, S3, S4 and S5), where results for different temperatures and applied loads are shown. 

\subsection{Machine learning}
To measure the performance of machine learning algorithms, usually statistical signatures like the coefficient of determination 
\begin{equation}
    R_{X}^2(t) = 1 - \frac{\sum_i(t_{\rm X,i} - t_{\rm a,i})^2}{\sum_i (\langle t_{\rm a,i} \rangle - t_{\rm a,i})^2}
    \label{r2}
\end{equation}
are used. Here the subscript X = ML refers to a machine learning prediction and X=BA to a baseline prediction. \(t_{\rm X,i}(t)\) is the predicted lifetime and \(t_{\rm a,i}\) the actual lifetime. The average $\langle .. \rangle$ is, in the following, always evaluated over the entire data set.

For failure time prediction it is important to know how the prediction quality evolves over time. Thus, we define time dependent prediction error and score:
\begin{equation}
    e_{\rm X}(t) = \frac{|t_{\rm X}(t) - t_{\rm a}(t)|}{t_{\rm f}}
    \label{error_eq}
\end{equation}
\begin{equation}
    s_{\rm ML}=1-\frac{e_{ML}}{e_{BA}}
    \label{score_eq}
\end{equation}
 were \(t\) is the time at which the prediction is made, and times are normalized by  the mean sample lifetime as  \(t_{f}\) normalization. The ML score should ideally be one or at least above zero which indicates superiority of the forest regression compared to the baseline estimate. These quantities are collected for each individual sample and sorted into bins according to the value of \(\frac{t}{t_{f}}\). For binning we use equi-spaced partitions which divide the unit interval into 100 bins for the FEM and 20 bins for the RFM data, thus accounting for the fact that the RFM produce fewer data points in a simulation.

\subsubsection{Baseline Evaluation}

\begin{figure}[thb]
\centering
\includegraphics[width=.9\textwidth]{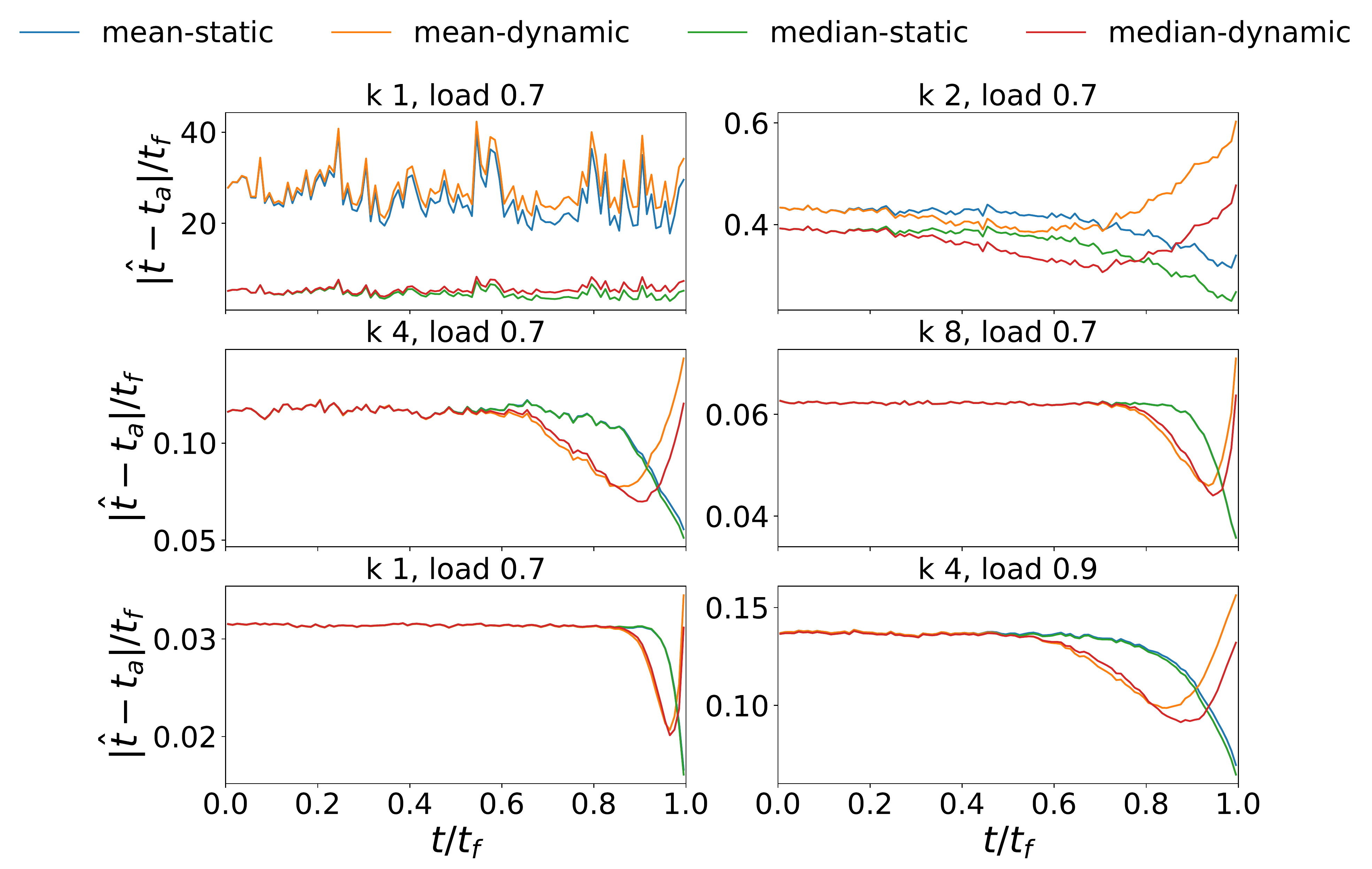}
\caption{Prediction error of different baselines, as function of time-to-failure in the FEM simulated data. The dynamic baselines calculate the mean or median lifetimes from the sub-population of the training set that still survives at time \(t\), whereas static baselines consider the entire population.}
\label{baseline-fem-errorplot}
\end{figure}

For the FEM data (Fig. \ref{baseline-fem-errorplot}) in the limit of high disorder, the mean lifetime performs very badly as a prediction of the lifetime of individual samples, irrespective whether the mean is taken over the total or the surviving population. The reason is the extraordinary high scatter of the lifetime distribution which exhibits a coefficient of variation much larger than one. The median lifetime, which is less dominated by extremely long-lived samples, performs better but still badly (Fig. \ref{baseline-fem-errorplot}, top left). 

When considering more ordered systems, the difference between median-based baselines and mean-based baselines vanishes especially at the start of the simulation up roughly half of the lifetime. This is expected as with decreasing disorder the lifetime distributions become less skewed, thus median and mean start to coincide. When close to failure, paradoxically, the dynamic baselines that consider only the surviving population perform worse than the static baselines that consider the whole initial population. This is the case because the surviving population exhibits a stronger outlier sensitivity: Its mean and median are increasingly dominated by the fittest samples which may show atypically high lifetimes.

\begin{figure}[ht]
\centering
\includegraphics[width=.8\textwidth]{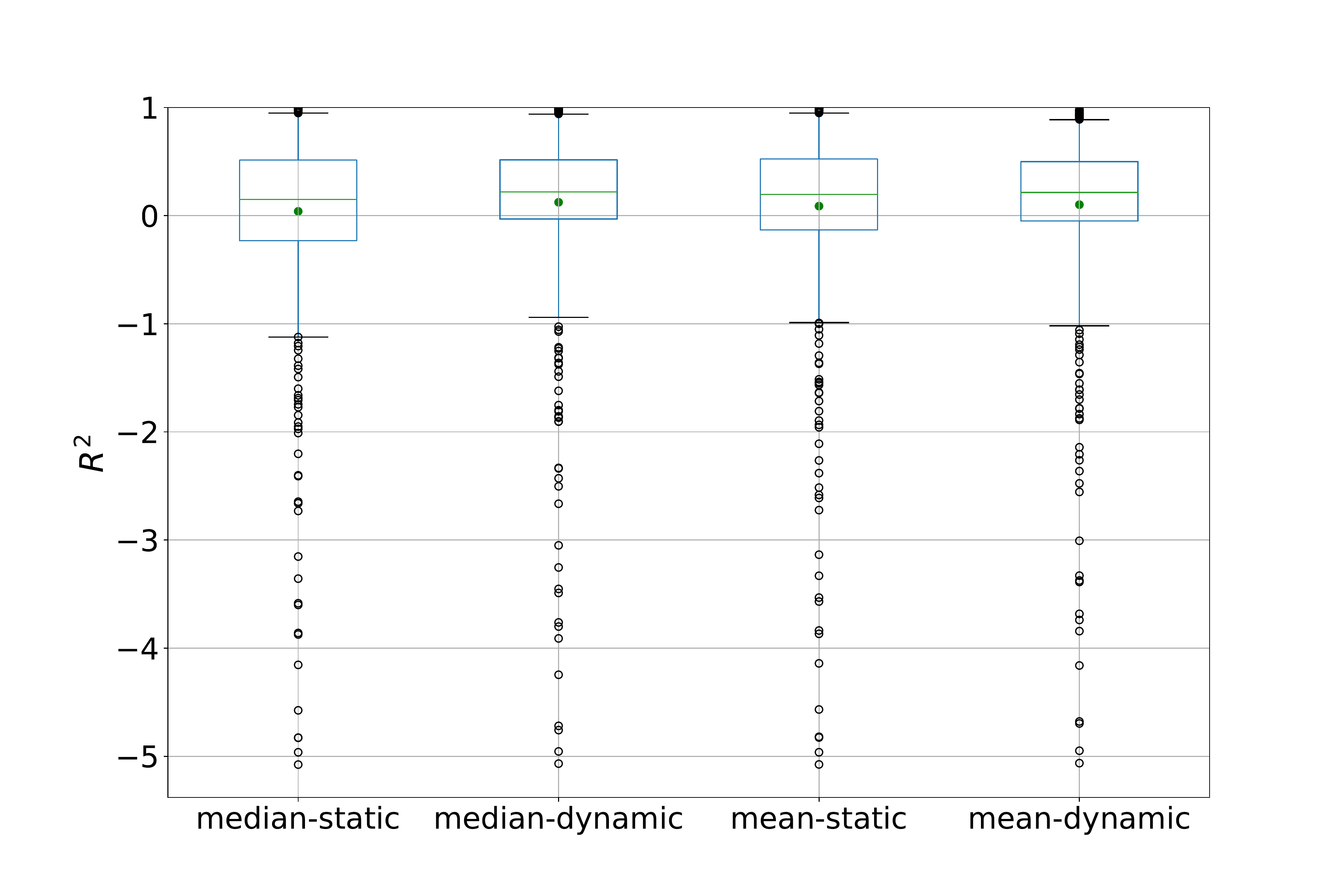}
\caption{Box and whisker plot of  \(R^{2}\) for RFM data with respect to different baseline choices. The data points correspond to different sets of simulation parameters, the green line indicates the median
\(R^{2}\), the green dot the average, the box bounds are located at the 25th and 75th percentile respectively and the whisker at the 5th and 95th percentile.}
\label{baseline-rfm-boxplot}
\end{figure}

To judge baseline performance across parameters, a box whisker plot (\ref{baseline-rfm-boxplot}) of \(R^{2}\) is used. Each data point in the plot represents the baseline performance for one combination of parameters (e. g. \(k=1\),\(T=0.01\),etc.). All baseline medians and averages (green line and dot) are above \(R^{2}=0\). The worst case performance assessed by the farthest outlier in the negative domain, for all four baselines is on the same level. The dynamically updated median has the highest average (green dot) and highest median performance(green line). The spread in performance between the dynamic baselines is similar, whereas the static baselines show a larger spread in performance (box length).

The static median is the best performing baseline for FEM, thus we choose it as baseline there. For RFM, the dynamic median is chosen as it has the highest average and median performance, although it is very similar to the dynamic mean.

\subsubsection{Performance}
\begin{figure}[ht]
\centering
\includegraphics[width=.8\textwidth]{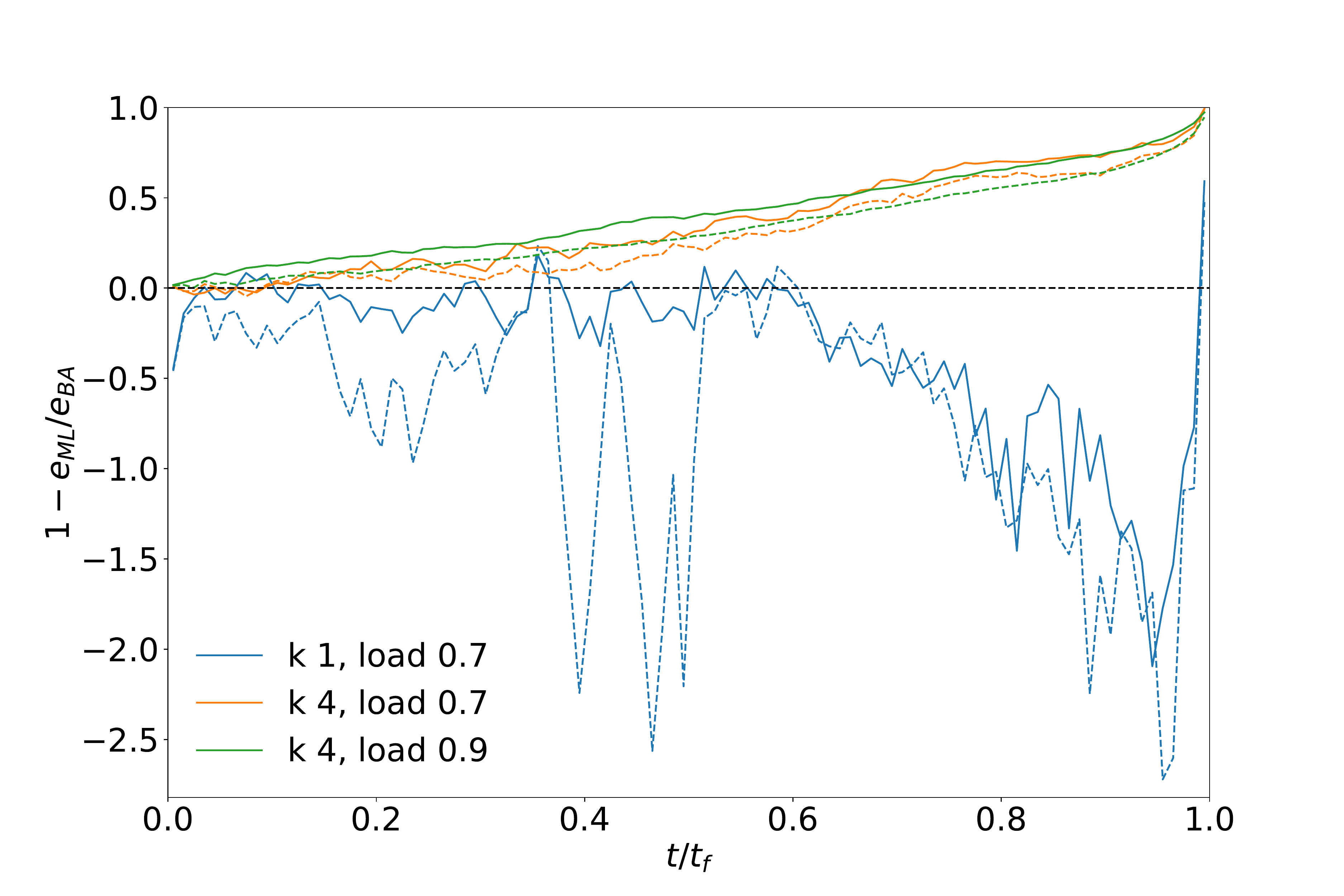}
\caption{Prediction score for the FEM data with the dotted lines representing feature set 1 already used in a previous publication and the full lines the new set 2. The baseline is the static median.}
\label{score-plots-fem}
\end{figure}

\begin{figure}[ht]
\centering
\includegraphics[width=.9\textwidth]{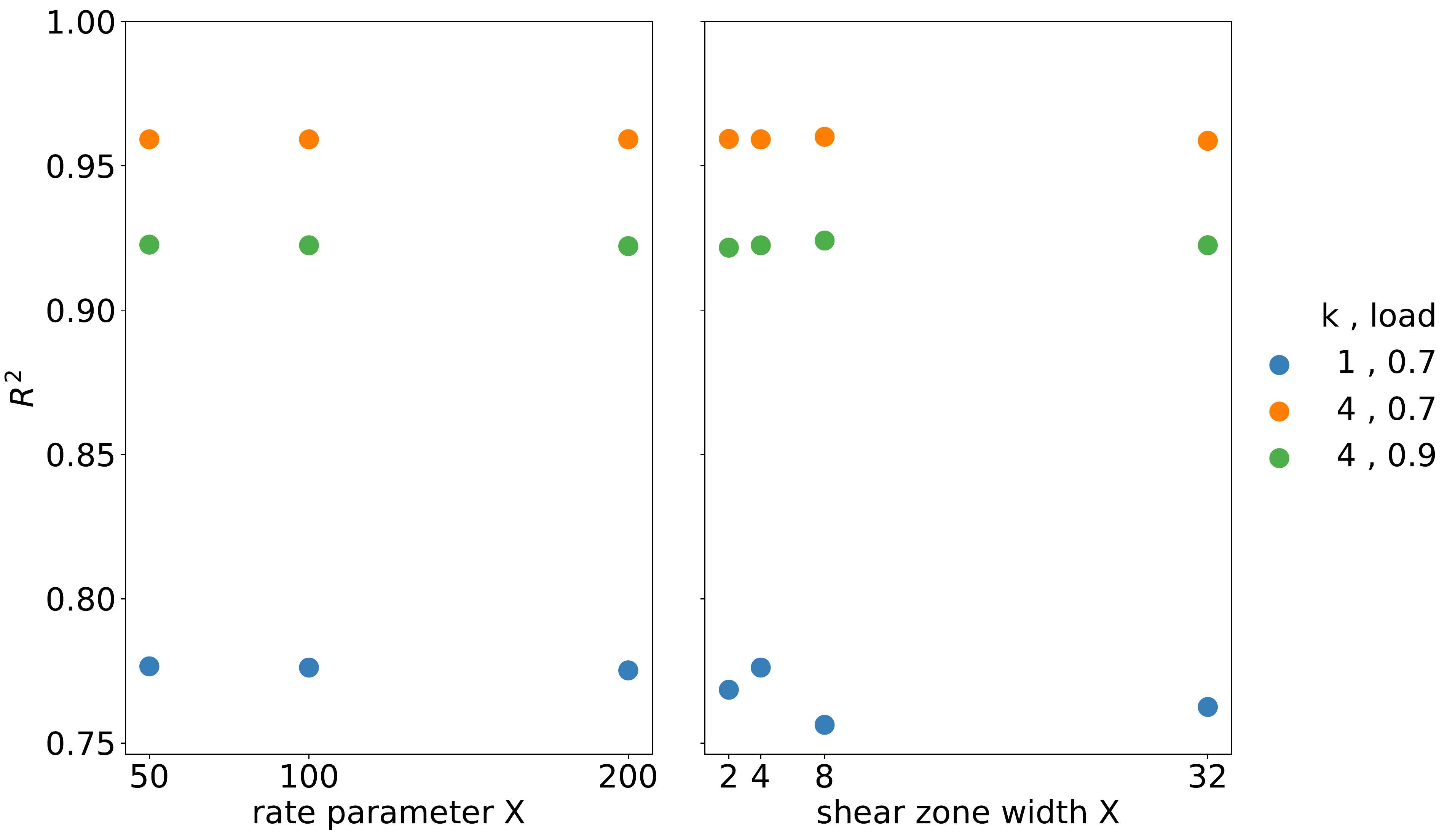}
\caption{Hyperparameters of feature set 2 for FEM data which show no significant impact on the prediction performance as measured by \(R^{2}\).}
\label{fem_parameter}
\end{figure}

When considering the score-time curve for the FEM data in Fig \ref{score-plots-fem}, the new feature set 2 (full lines) performs slightly better than the feature set used by Biswas et. al.  \cite{biswas}, but for the highly disordered case, both feature sets fail to outperform the baseline until just before the failure event. Even then, the mean error is still of the order of five times the mean residual lifetime, thus the prediction cannot be called successful. For less disordered samples, the ML algorithm outperforms the baseline across the entire specimen lifetime. The time series parameter \(X\)) does not affect the performance while the shear zone width shows a small peak performance at a width of four elements, but not a game changing improvement (Fig. \ref{fem_parameter}). As established in previous work \cite{biswas}, different applied loads also do not strongly affect the quality of the ML model predictions (Fig. \ref{score-plots-fem}) even though the absolute lifetimes may change by many orders of magnitude. Higher disorder generally leads to a decrease in prediction performance. 

\begin{figure}[htb]
\centering
\includegraphics[width=\textwidth]{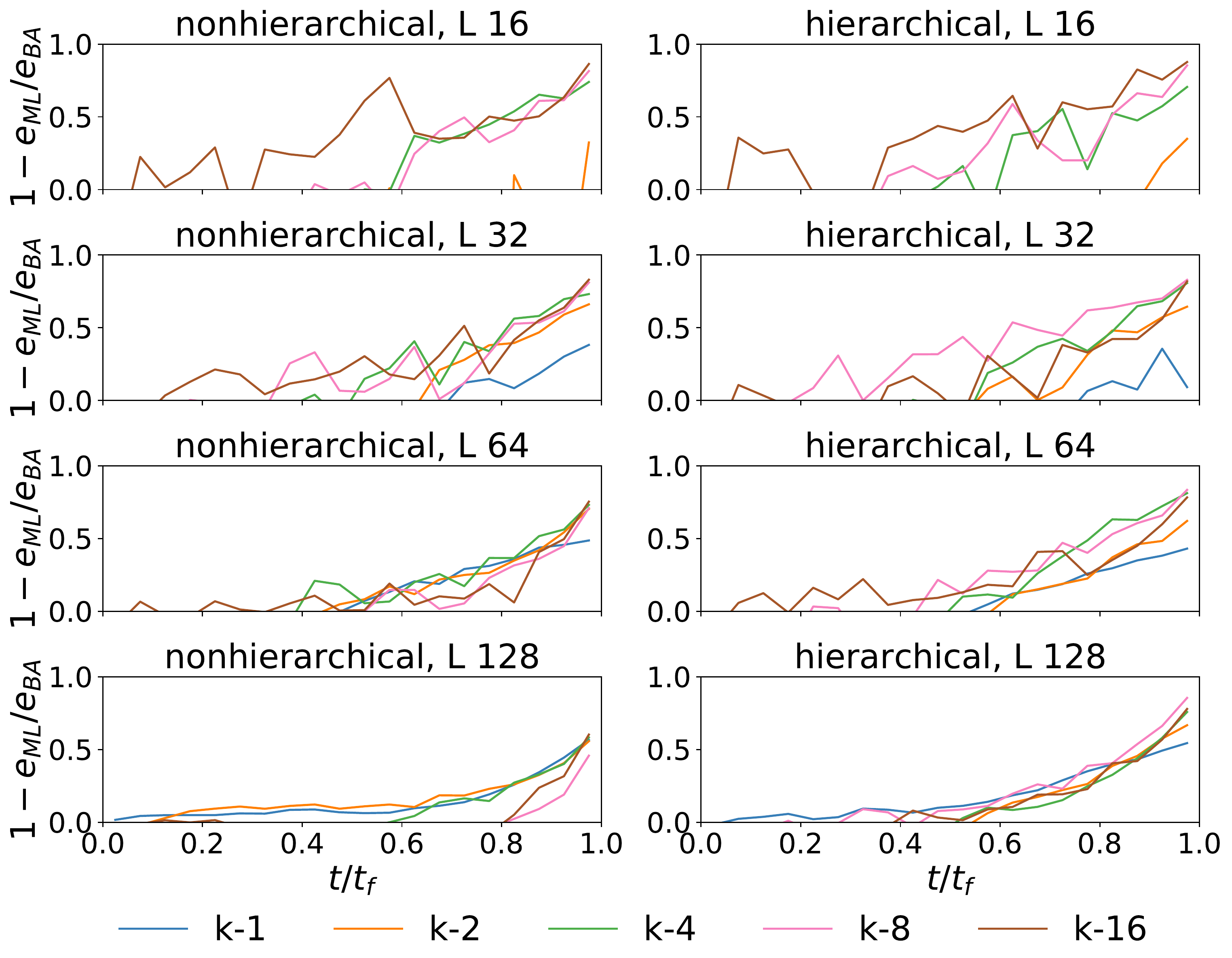}
 \caption{Prediction score calculated for system with \(T=0.03\) and \(\frac{I}{I_{p}}=0.7\) for different values of \(k\). The ML models were trained on feature set 1. The baseline for comparison is the dynamic median.}
\label{time-score-k-1}
\end{figure}

\begin{figure}[ht]
\centering
\includegraphics[width=\textwidth]{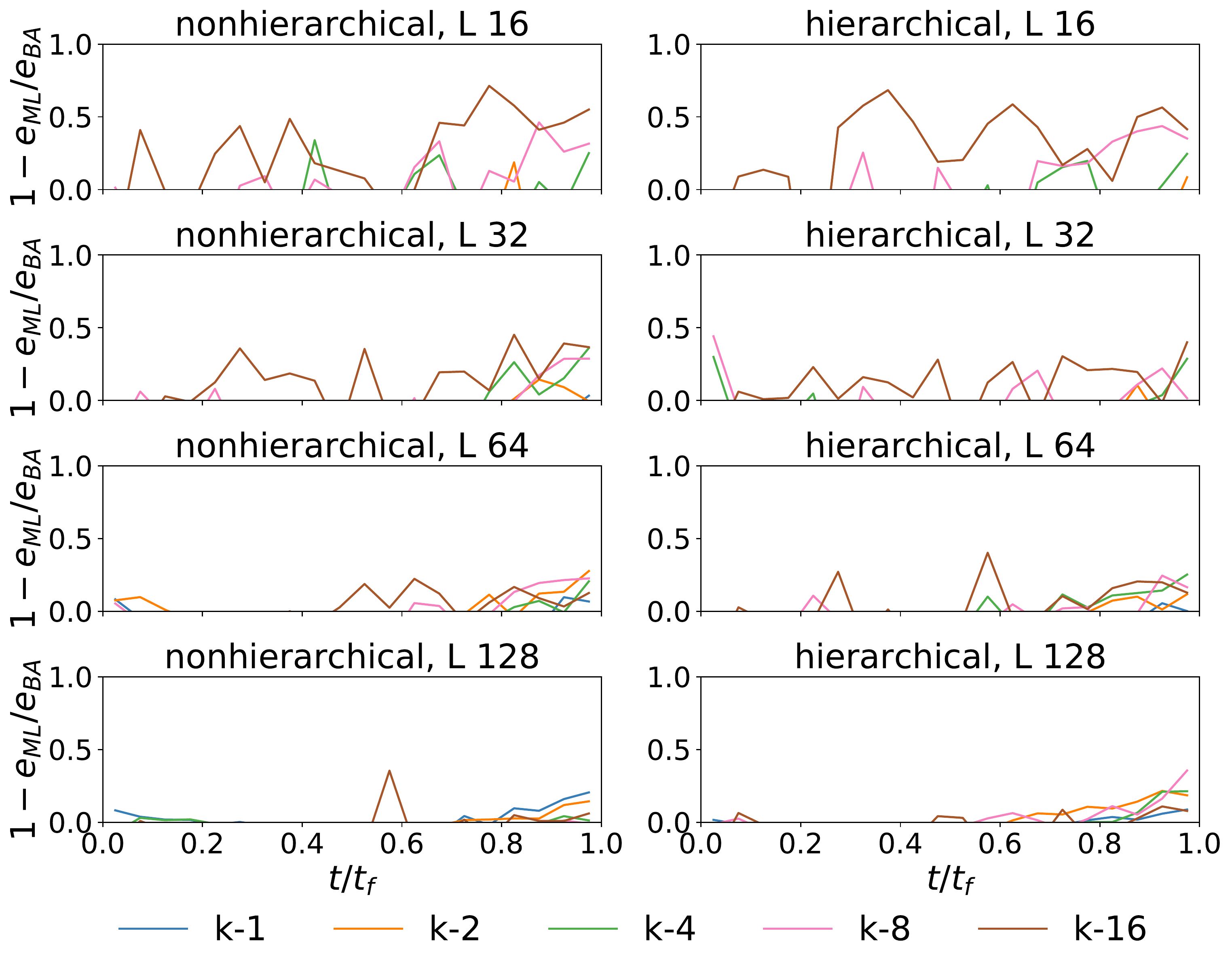}
\caption{Prediction score calculated for system with \(T=0.03\) and \(\frac{I}{I_{p}}=0.9\). The ML models were trained on feature set 2. The baseline for comparison is the dynamic median.}
\label{time-score-k-2}
\end{figure}

Random forests trained on RFM data can outperform the corresponding dynamic median baseline, but it depends strongly on the system parameters to which extent this is the case as can be seen when comparing Figures \ref{time-score-k-1} and \ref{time-score-k-2}. Small systems show a less stable trend as these systems produce only few thermally activated events before failure and thus deliver less reliable statistics for the plot, but also make the task harder for the random forest. Similar problems are encountered in ordered systems where, owing to the small scatter in strength, load re-distribution produces large avalanches, resulting again in a small number of (though large) events (Fig. S5) and poor statistics. 
\begin{table}[ht]
\caption{Average rank correlation between the prediction performance \(R^{2}\) and the simulation parameter for RFM data. Correlations from random forests trained on feature set 1 are at the upper part of the table, the same for feature set 2 on the lower part.}
\centering
\begin{tabular}{ccccc}
\hline
parameter & nonh. kendall & nonh. spearman & hier. kendall & hier. spearman \\ 
\hline
 size & 0.247 & 0.234 & 0.283 & 0.267 \\ 
k & 0.088 & 0.062 & 0.412 & 0.469 \\ 
temperature & 0.85 & 0.88 & 0.35 & 0.33 \\ 
current & -0.683 & -0.72 & -0.567 & -0.59 \\
\hline
size & 0.247 & 0.226 & 0.27 & 0.245 \\ 
k & 0.062 & 0.05 & 0.425 & 0.444 \\ 
temperature & 0.8 & 0.83 & 0.433 & 0.43 \\ 
current & -0.667 & -0.71 & -0.6 & -0.6 \\
\hline
\end{tabular}
\label{rank-correlation-r2-table}
\end{table}

To gain more systematic insight how ML performance depends on the simulation parameters, we calculate the average rank correlation as illustrated in the following for the system size: We iterate over all combinations of  \(k\),\(T\) and \(\frac{I}{I_{p}}\), calculate for each combination the rank correlation between system size \(L\) and \(R^{2}\) and take the average over all combinations. The results are compiled in Table \ref{rank-correlation-r2-table}. 

Increasing temperature,system size and order lead to increased prediction performance while increasing current does the opposite, but for disorder and temperature the correlation depends strongly on the network architecture. The trend with regards to temperature is most stable for the nonhierarchical networks. In the low-temperature limit, the crack path is identical to the quasistatic extremal fracture path whereas increasing temperature leads to activity at other sites of the network which in turn yields information for the random forest, therefor better predictions. In the hierarchical network load redistribution is already known to lead to activity spread across the system contrary to the nonhierarchical case \cite{moretti2018avalanche}, thus activity spread through thermally activated failure might yield less additional information.
High currents and small systems lead to quick failure, thus less information to work with. The effect of disorder is ambiguous as high disorder leads to an increasing range of possible fracture paths thus a larger feature space, but also to more activity as a source of information, while low disorder can lead to fracture after few thermally activated events, creating little information about fracture precursors that might be used to improve predictions above the baseline. 

\subsubsection{Feature Selection}

\begin{figure}[htb]
\includegraphics[width=\textwidth]{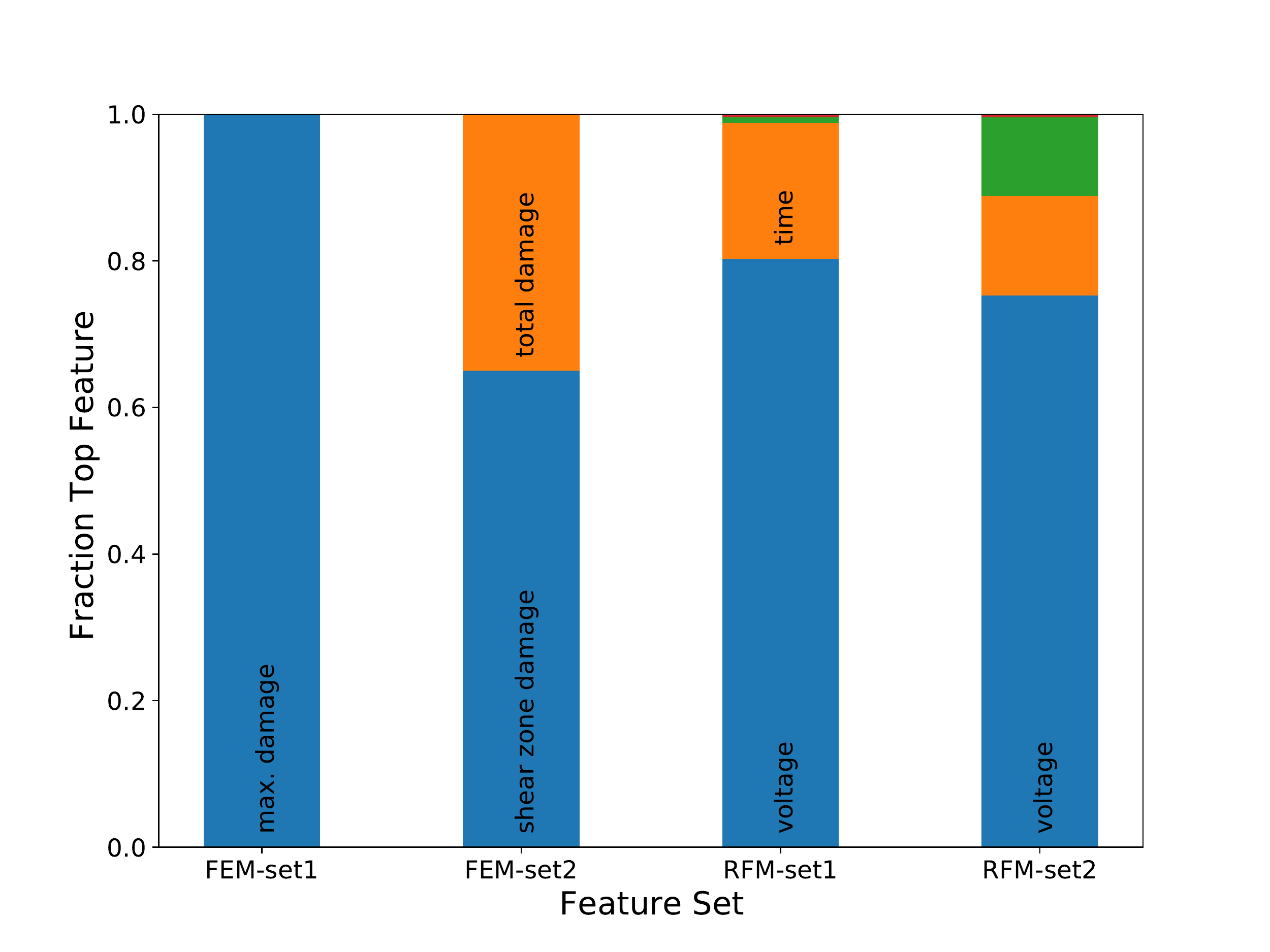}
\caption{Fraction of times a feature has been the most important one in a model trained on the specified feature set.}
\label{Feature-count}
\end{figure}

Regarding feature selection two main questions should be answered: What are the most important features and how focused is the random forest on the top features? If the model attention is spread equally among all features, then no reduction in dimensionality is possible.
For each feature set, we count how often a specific feature is deemed the most important according to the normalized average error reduction criterion (Fig. \ref{Feature-count}). To measure the attention spread of the model, we calculate the Gini Impurity \cite{breimanforest}
\begin{equation}
    1 - \sum_{i=1}^{N_{\rm feat}}p_{i}^{2}
\end{equation}
from the normalized average error reduction (aer) \(p_{i}\) (\(\sum_{i=1}^{N_{\rm feat}}p_{i}=1\)) of feature \(i\).  \(N_{\rm feat}\) is the number of features of the model. Close to zero Gini impurity indicates a model focused on a single feature. 

Fig. \ref{Feature-count} shows that damage based features dominate the partitioning process for FEM data whereas models for predicting RFM behavior mainly use voltage (i.e., strain) as the source of their predictions. The importance of damage based features for the FEM model is easily rationalized, as this model shows shear banding behavior which leads to a large total damage as well as high damage concentration in the incipient shear band. 
\begin{table}[ht]
\caption{Average rank correlation between different simulation parameters and the Gini impurity of the ML model, for RFM data. Correlations from models trained on feature set 1 are at the upper part of the table, correlations for feature set 2 at the lower part.}
\center
\begin{tabular}{ccccc}
\hline
parameter & nonh. kendall & nonh. spearman & hier. kendall & hier. speaman \\ 
\hline
size & -0.53 & -0.536 & -0.423 & -0.419 \\ 
k & -0.338 & -0.375 & -0.587 & -0.706 \\ 
temperature & -0.95 & -0.97 & -0.9 & -0.94 \\ 
current & 0.75 & 0.74 & 0.45 & 0.44 \\ 
\hline
size & -0.507 & -0.519 & -0.52 & -0.524 \\ 
k & -0.55 & -0.625 & -0.488 & -0.569 \\ 
temperature & -0.967 & -0.98 & -0.717 & -0.74 \\ 
current & 0.667 & 0.67 & 0.583 & 0.62 \\ 
\hline
\end{tabular}
\label{rank-correlation-gini-table}
\end{table}

For feature set 1 the avalanche amplitude is the least important feature regardless of investigated system and with a normalized aer ranging from zero to \(p = 5.71\times10^{-6}\). In feature set 2 the event rate and averaged avalanche amplitude share the least important feature spot which never exceeds an aer of \(p = 1.29\times10^{-4}\). In fact, if the event rate is the least important feature, the second least important feature is always the averaged avalanche amplitude which never exceeds an aer of \(2.56\times10^{-2}\). It is thus safe to say that, for our FEM data and feature sets, the avalanche amplitude and rate are not very useful for failure time prediction.

\begin{figure}[p] 
\includegraphics[width=\textwidth]{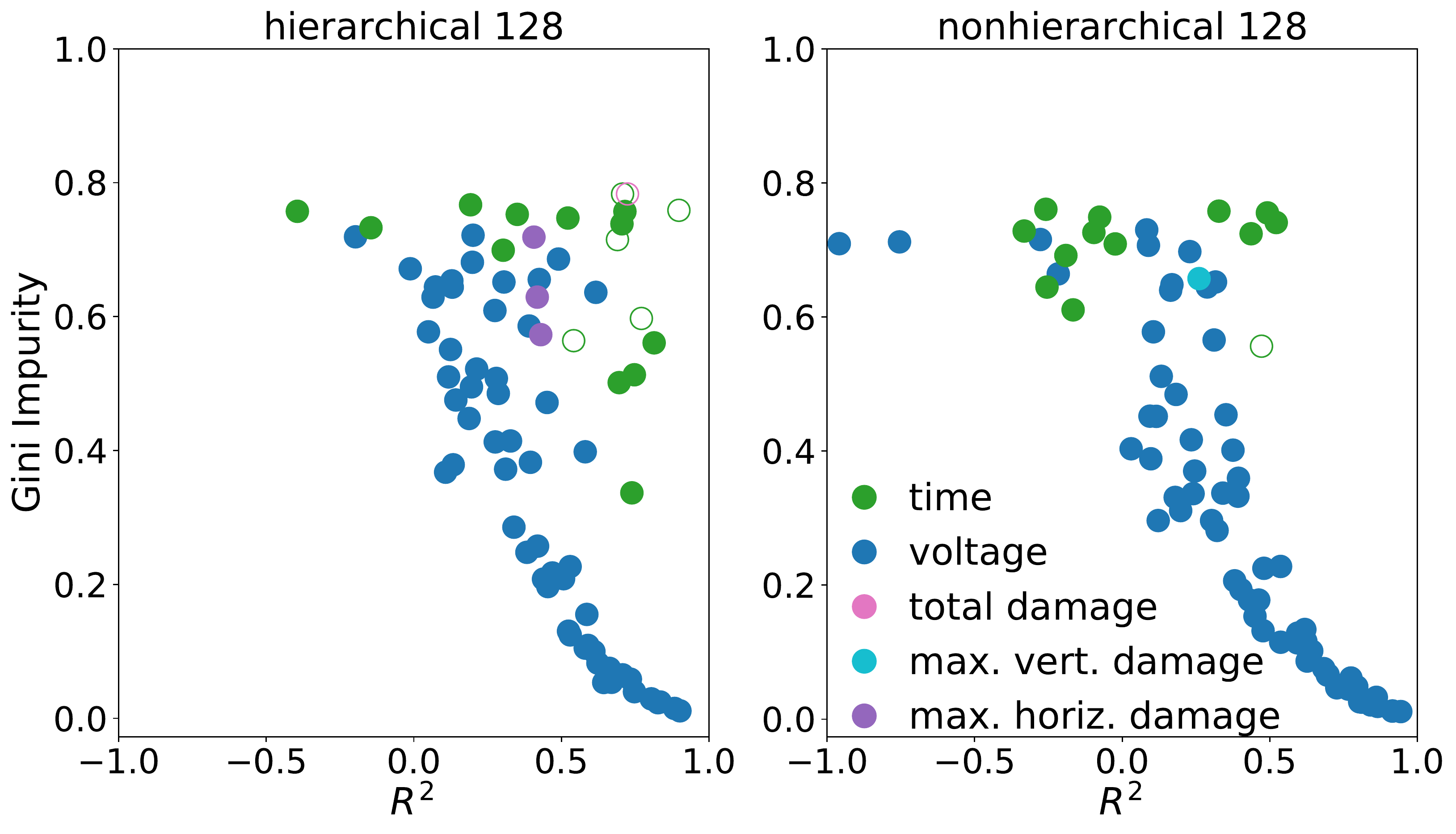}
\includegraphics[width=\textwidth]{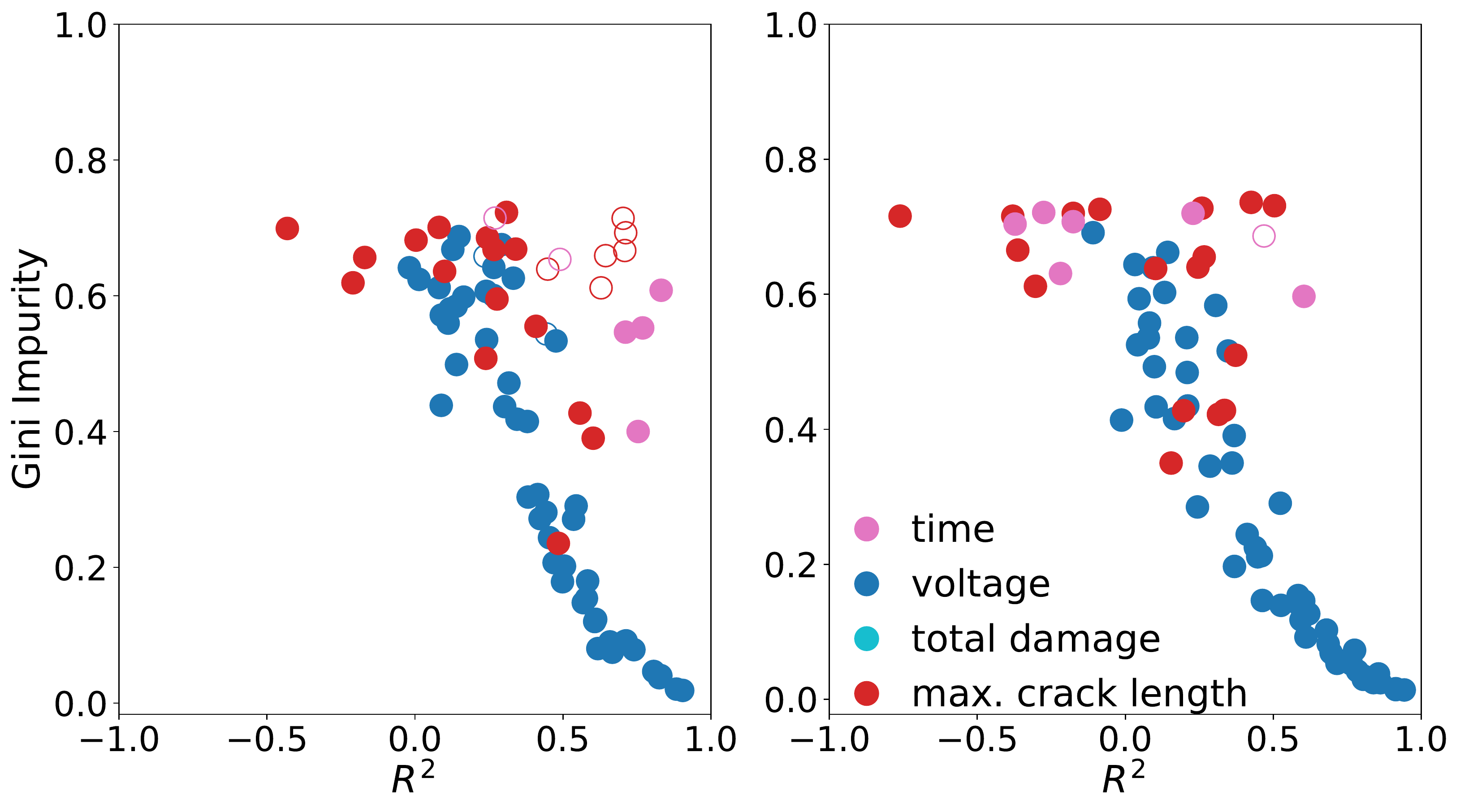}
\caption{Gini Impurity as a measure of attention spread versus model performance of the RFM data with hierarchical network data on the left and nonhierarchical on the right. The top row stem from ML models trained on feature set 1 and the bottom row on feature set 2. The color shows the top feature for the specific data point. Hollow dots indicate that for this data point the static median baseline's \(R^{2}\) is higher than the random forest model.}
\label{gini-r2}
\end{figure}

For the RFM data, models with high attention spread can occur where even the least important feature has an aer of \(p \approx 0.17\). To investigate the consequences, we compare the model performance as measured by \(R^{2}\) with the Gini impurity in Fig. \ref{gini-r2}. Every data point represents one combination of model parameters and the color represents the most important feature of the model trained on the specific combination. Attention focus and model performance correlate positively, although this correlation is less pronounced for hierarchical networks where in a number of cases, attention diversified models achieve comparable performance to focused models. When we calculate average rank correlations between the Gini impurity and the model parameters in the same ways as we have done previously for \(R^{2}\) (Table \ref{rank-correlation-gini-table}), we can see that the numerical values of temperature and system size correlate positively with attention concentration, while high current/load and high disorder have the opposite effect. This trend is the most stable for temperature with some deviations in the case of feature set 2 and hierarchical networks. When comparing Table \ref{rank-correlation-r2-table} and \ref{rank-correlation-gini-table}, we notice that the different simulation parameters tend to have opposite-sign correlations with $R^2$ and Gini impurity, further supporting the connection between \(R^{2}\) and ML model attention.

\section{Conclusion}

In this work, we determine the predictive value of a set of spatial and temporal features  to predict the time to failure in creep simulations by machine learning. The data for this study stem from a random fuse model (RFM) describing brittle, thermally activated statistical failure and an elasto-plastic finite element method (FEM) approach \cite{castellanos2018} which has been used in a previous study \cite{biswas}. Fuse networks are created with two different morphologies (hierarchical vs nonhierarchical) as hierarchical architecture was shown to modify systems behavior in the approach to failure \cite{moretti2018avalanche}. To measure performance, we compare the trained models with a simple statistical baseline. We make use of the ability of a special method of data science (random forest/decision trees) to measure the importance of a single feature for predictions via the average error reduction. Predictability of RFM systems is found to increase with increasing size, temperature and order, whereas increasing load/current has the opposite effect. 
The features deemed most important by the random forest algorithm for prediction of FEM simulation data are the damage localized in a shear band and the total damage,  whereas predictions for RFM data rely mainly on voltage. Note that  voltage is the equivalent of strain in the RFM model and that damage in the FEM model is a monotonically increasing function of strain. Avalanche-related features and event rate were found to be of no predictive power for the case of FEM, whereas RFMs are too brittle -- there are not enough events to calculate a sensible time dependent event rate. This suggests that future experimental studies - contrary to existing literature suggesting strain rate \cite{hao2014predicting,koivisto2016predicting,nechad2005creep}, event rate and amplitude \cite{castellanos2018,lennartz2014acceleration,saichev2005andrade} - should consider the strain as a possible valuable feature for the predictions of the time to failure in a creep setting. Note that spatial patterns of strain might, in real life situations, be accessed nondestructively by surface monitoring. 

\section*{Acknowledgments}
S.H., M.Z. and S.Z. acknowledge support from the Deutsche Forschungsgemeinschaft (DFG) under Grant no. ZA171/14-1. S.H. acknowledges participation in the training programme of the DFG graduate school FRASCAL, 377472739/GRK 2423/1-2019.

\section*{Competing interests}
The authors declare no competing interests.

\bibliographystyle{elsarticle-num} 
\bibliography{cas-refs}






\end{document}


\maketitle 
\newpage

\section{Quasistatic Loading, RFM} 
\begin{figure}[ht]
\centering
\includegraphics[width=\textwidth]{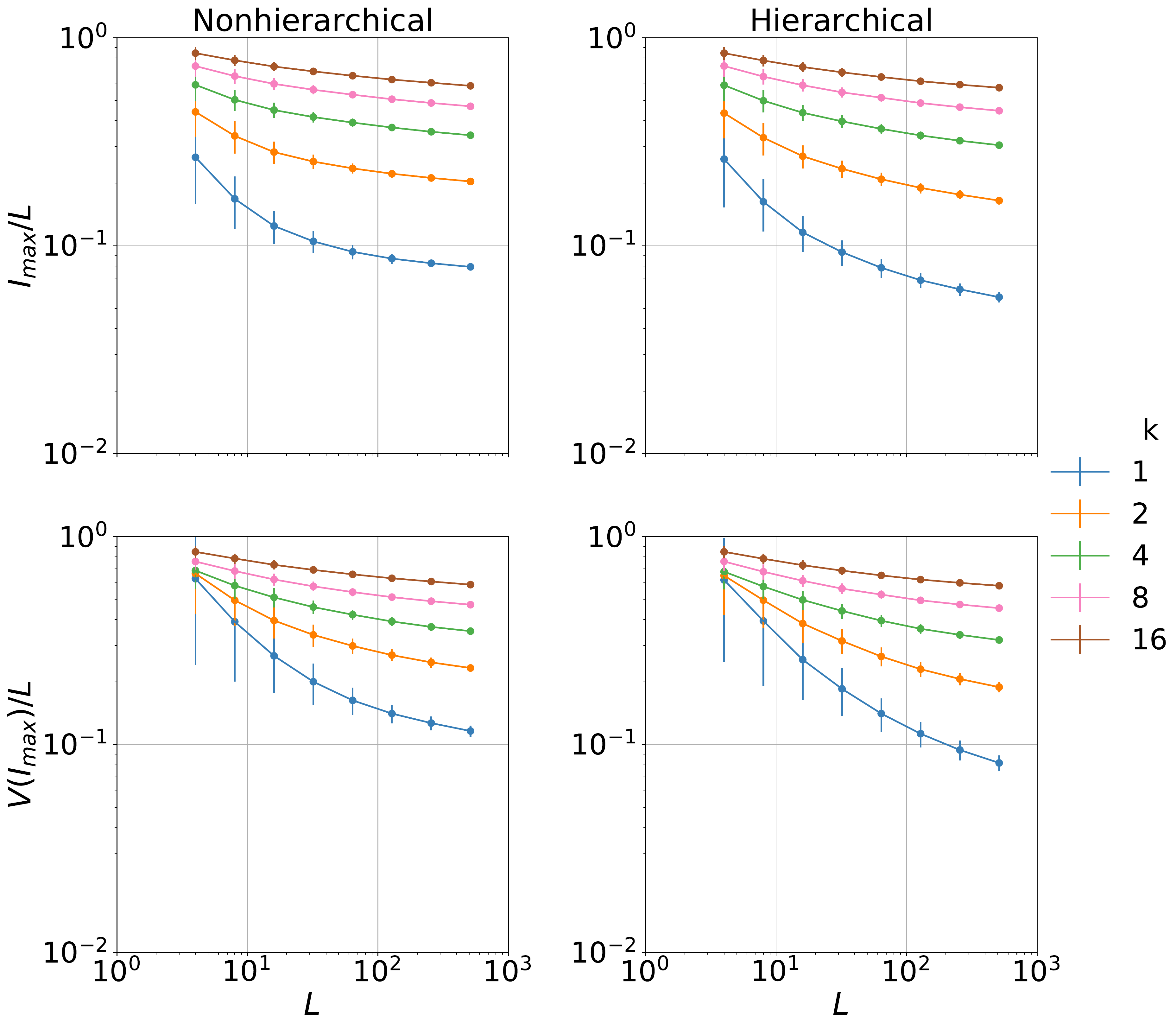}
\caption{Average maximum current and average voltage at maximum current for the RFM model, as functions of system size and disorder.}
\label{mean-quasistatic}
\end{figure}
\newpage
\section{RFM creep data: dependency of creep curves on simulation parameters}
\begin{figure}[ht]
\includegraphics[width=\textwidth]{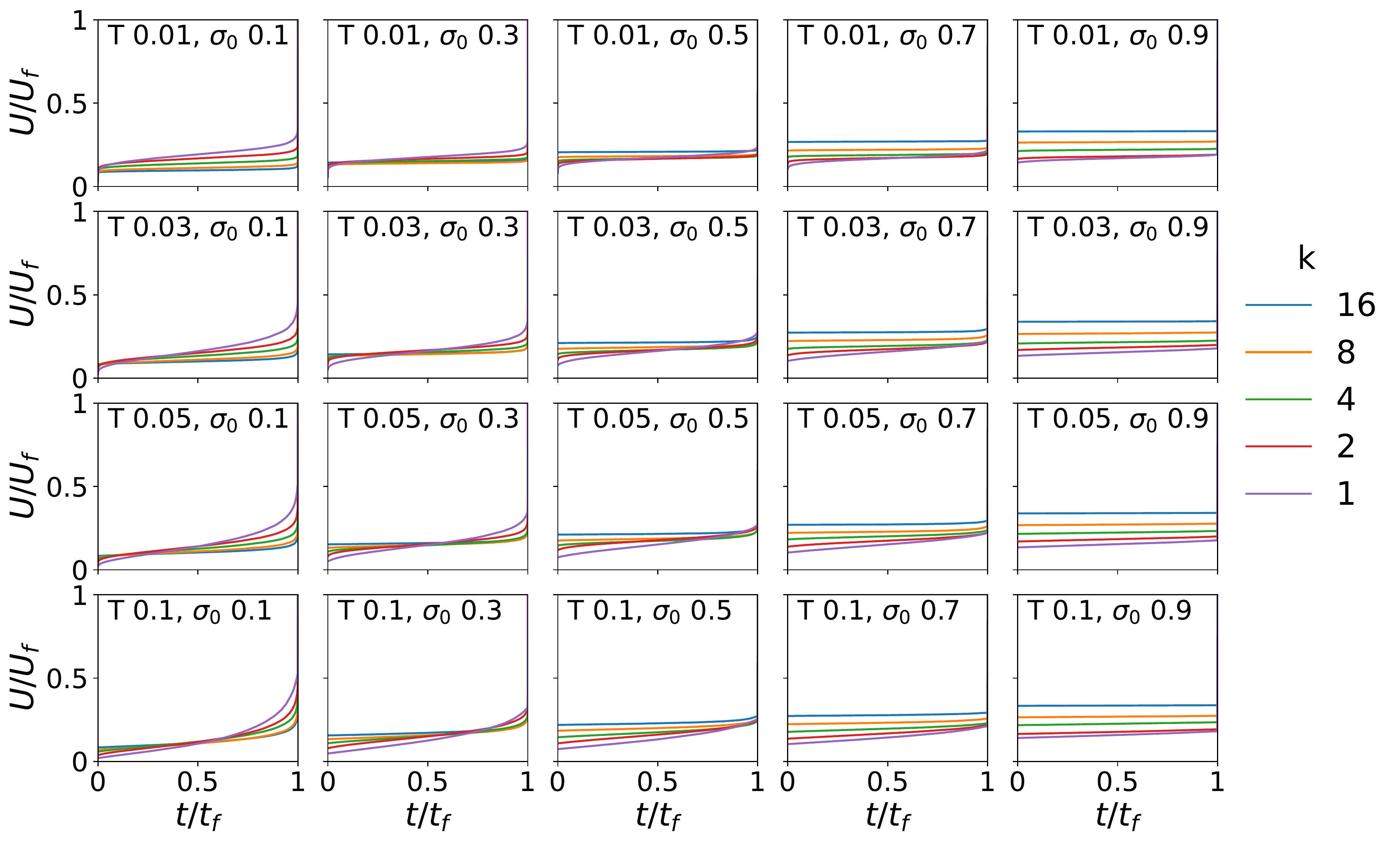}
\caption{Normalized time versus normalized voltage for hierarchical networks of size \(L=128\) and various simulation parameters.}
\label{t-U-hierarchical}
\end{figure}
\begin{figure}[ht]
\centering
\includegraphics[width=\textwidth]{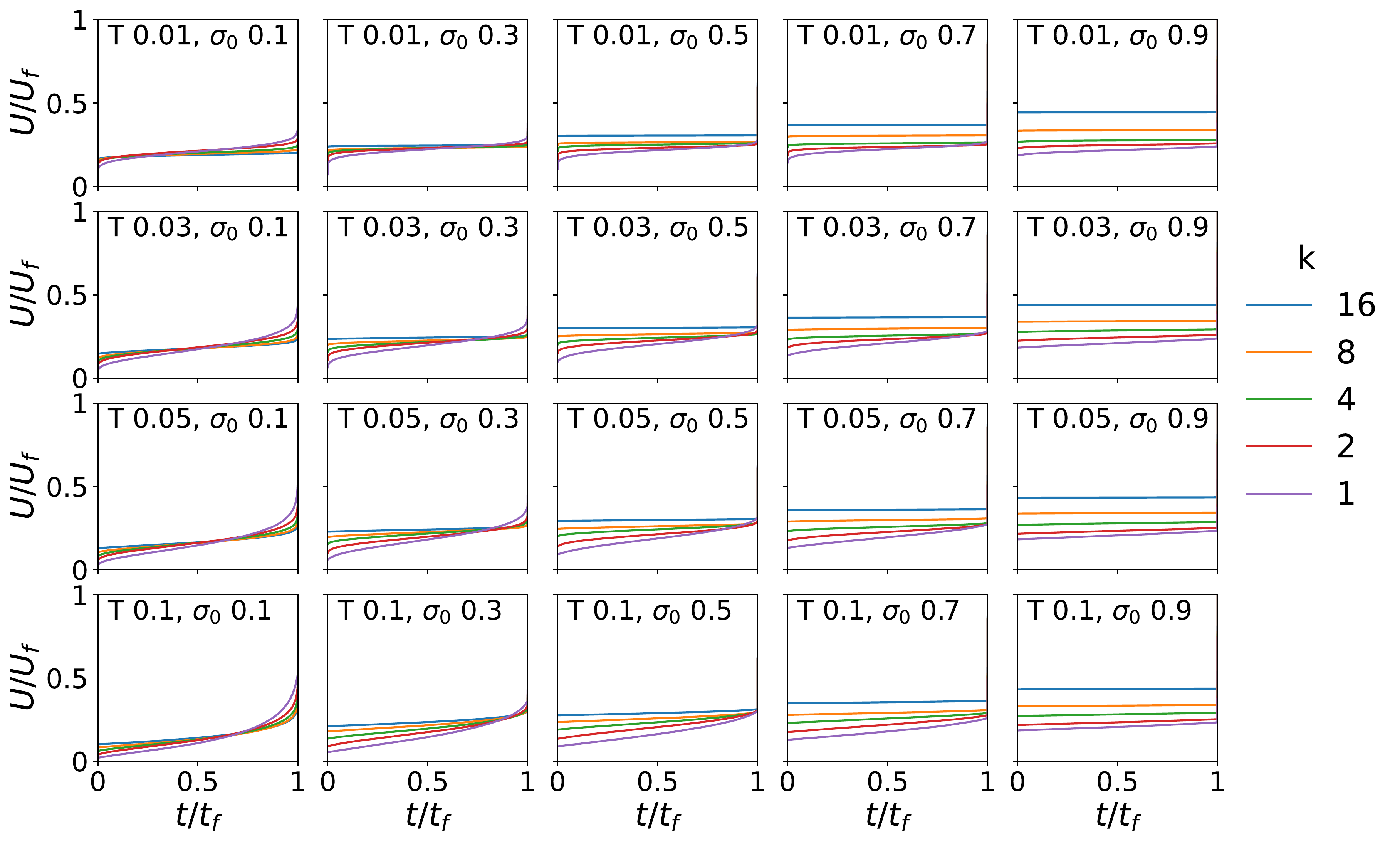}
\caption{Normalized time versus normalized voltage curves for nonhierarchical networks of size \(L=128\) and various simulation parameters.}
\label{t-U-nonhierarchical}
\end{figure}
\begin{figure}[ht]
\centering
\includegraphics[width=\textwidth]{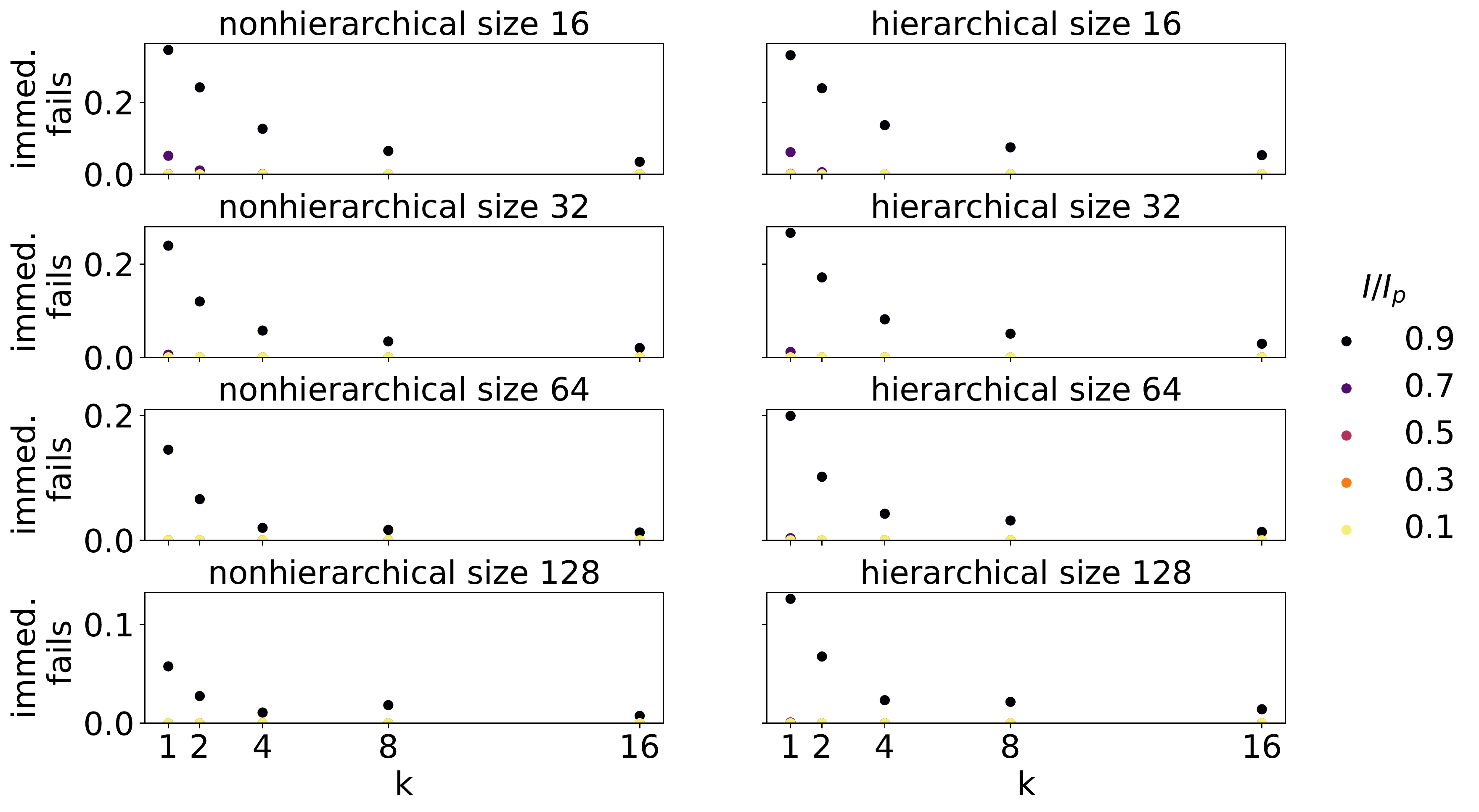}
\caption{Percentage of networks failing immediately at time zero in RFM simulations, for different simulation parameters.}
\label{immed. failure}
\end{figure}
\begin{figure}[ht]
\centering
\includegraphics[width=\textwidth]{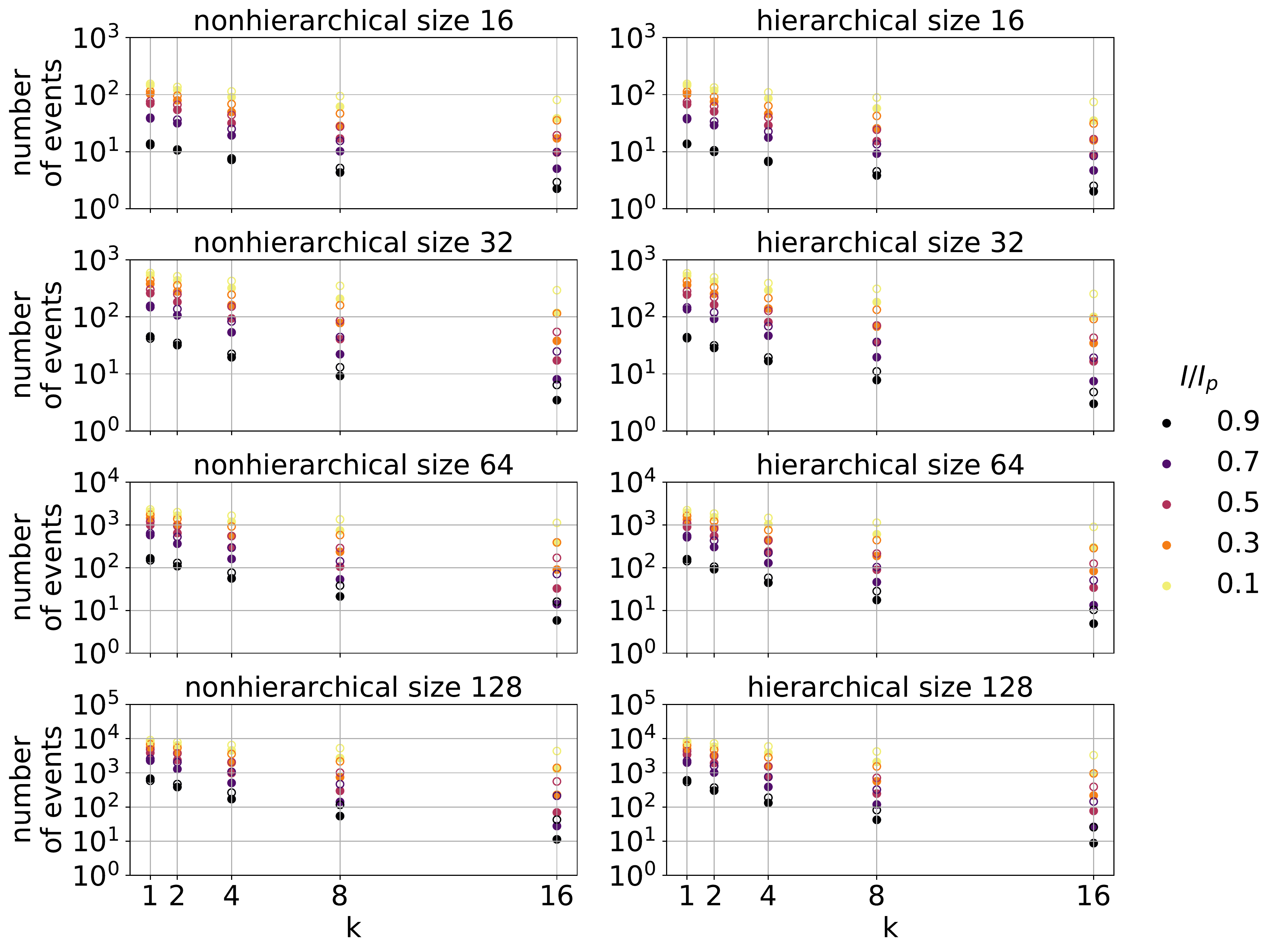}
\caption{Number of events before failure in RFM simulations, for different simulation parameters. Full symbols: temperature 0.01, empty symbols: temperature 0.1.}
\label{nevents}
\end{figure}  


